\documentstyle[version2,aps]{revtex}
\begin{document}
\draft
\columnsep -.375in
\twocolumn[
\begin{title}
Quantization of Superflow Circulation and
Magnetic Flux with a Tunable Offset
\end{title}
 \author{
 Yuli Lyanda-Geller\rlap,\cite{YGnote}$^{(a)}$
 Paul M.~Goldbart\cite{PGnote}$^{(a,b,c)}$ and
 Daniel Loss\cite{DLnote}$^{(d)}$}
\begin{instit}
$^{(a)}$Beckman Institute,
$^{(b)}$Department of Physics and
$^{(c)}$Materials Research Laboratory,\\
University of Illinois at Urbana-Champaign,
Urbana, Illinois 61801, USA;\\
$^{(d)}$Department of Physics,
Simon Fraser University,
Burnaby, B.~C.~V5A 1S6, Canada
\end{instit}
%
\date{May 29, 1995}
\begin{abstract}
Quantization of superflow-circulation and of magnetic-flux are
considered for systems, such as superfluid $^3$He-A and unconventional
superconductors, having nonscalar order parameters.  The circulation is
shown to be the anholonomy in the parallel transport of the order
parameter.  For multiply-connected samples free of distributed
vorticity,  circulation and flux are  predicted to be quantized, but
generically to nonintegral values that are tunably offset from
integers. This amounts to a version of Aharonov-Bohm physics.
Experimental settings for testing these issues are discussed.
\end{abstract}
\pacs{PACS numbers:
03.65.Bz,	
67.57.-z,	
74.20.-z 	
\hfil\break
Preprint Number: P-95-05-036-ii
}
]
\narrowtext
\newpage
\noindent{\it Introduction\/}:
In superfluid $^4$He, as well as in conventional superconductors, for
{\it any\/} configuration of the
%
	appropriate order parameter,
%
and for {\it any\/} closed path through nonsingular regions of it, the
superflow circulation $\kappa$ adopts one of only a discrete set of
values.  Quantum-mechanical in origin \cite{REF:lo_rpf_rd,REF:byandya},
this remarkable phenomenon is referred to as the quantization of
circulation.  In $^4$He, circulation occurs at integral multiples of
$\hbar/M$, $M$ being the mass of the $^4$He atom.  (In conventional
superconductors, $M$ is the Cooper-pair mass).\thinspace\ The magnetic
flux enclosed in a ring of conventional superconductor is also
quantized, in units of $\Phi_0$ ($=2.07\times 10^{-15}\,{\rm We}$).  In
superfluid $^3$He, however, $\kappa$ is not necessarily quantized,
owing to the anisotropy of certain of its superfluid phases
\cite{REF:review_ajl,REF:wph_lpp,REF:review_vw}.

For $^3$He-A, the standard view \cite{REF:review_vw} is that despite
the absence of circulation quantization in the {\it bulk\/}, integral
quantization, precisely as in $^4$He, occurs at {\it surfaces\/}, owing
to the anchoring of the relative orbital
%
	angular momentum vector ${\bf l}$.
%
In fact, we shall see that quantization at surfaces is by no means the
rule, instead being the exception, with integral quantization being yet
more exceptional. Indeed, the aim of this Letter is to demonstrate
that, although quantization is not generic, there do exist
experimentally achievable order parameter configurations for which
circulation is quantized, but not to conventional (i.e., integral)
values.  Instead, what occurs is {\it offset circulation quantization,
i.e., integrally-spaced but nonintegral circulations\/}:
$\cdots,-1+\omega,0+\omega,1+\omega,\cdots$, for all paths.
The offset $\omega$ can be continuously tuned by varying the sample
shape or certain external fields, both of which can determine the
texture of the order parameter.
Moreover, as we shall see,
	the Aharonov-Bohm--like
physics \cite{REF:aha_boh,REF:as_fw} responsible for offset
circulation-quantization in $^3$He-A also leads to the possibility that
a {\it nonintegral number of magnetic flux quanta\/} could be enclosed
in a ring of unconventional superconductor.  Furthermore, if the order
parameter is in equilibrium then offset quantization has the striking
corollary that arbitrary nonintegral equilibrium circulations can be
obtained, in contrast with the standard possibilities:
$0,1/2,1$ (bulk); $0,1/2,1,3/2,\ldots$ (surface);
see, e.g., \cite{REF:review_vw}.

We proceed as follows.  First, we establish a general relationship
between circulation, parallel transport and order parameter
anholonomy.  Next, we introduce Aharonov-Bohm--type concepts to this
area of physics.  Finally, we discuss experimental configurations for
superfluid $^3$He-A, and propose tests involving the offset of
magnetic-flux--quantization in unconventional superconductors and the
phase-shift of Little-Parks oscillations \cite{REF:LandP}.

\noindent{\it Geometric phase\/}:
The origin of the phenomena considered here is
%
	the geometric phase
%
\cite{REF:mvb_qap,REF:as_fw}, i.e., the phase acquired by Cooper pairs
as they propagate through the condensate that they self-consistently
form, their relative (orbital or spin) angular momentum following the
local orientation determined by the condensate.  This geometric phase
is an \lq\lq anholonomy:  the geometrical phenomenon in which
nonintegrability causes variables to fail to return to their original
values, when others, which drive them, are altered round a
cycle\rq\rq\ \cite{REF:mvb_qpfyl}.  Indeed, as we shall show, the
circulation computed around any path in an anisotropic superfluid is a
natural anholonomy in the parallel transport of the triads that
characterize its order parameter.

Consider superfluid $^3$He-A, which is characterized by an
%
	order parameter matrix
%
$d_{\mu j}$ of the form:
\begin{equation}
d_{\mu j}=
\Delta_{0}\,
d_{\mu}
\big(m_{j}+in_{j}\big)
\exp i\chi.
\end{equation}
Here, $\Delta_{0}$ is the (real) gap parameter for the condensate,
$d_{\mu}$ is a (real) unit-vector associated with the spin angular
momentum sector, $\{{\bf l},{\bf m},{\bf n}\}$ is an orthonormal,
right-handed, triad of (real) vectors associated with the relative
orbital angular momentum sector, and $\chi$ is a (real) phase.  This
parametrization of $d_{\mu j}$ in terms of $\chi$ and
$\{{\bf l},{\bf m},{\bf n}\}$ is not unique: $d_{\mu j}$ is invariant
under the simultaneous operations of rephasing $\chi$ (by $\delta\chi$)
and rotating $\{{\bf m},{\bf n}\}$ about ${\bf l}$ (by $-\delta\chi$).
Due to the broken relative gauge-orbital symmetry (for reviews, see
Refs.~\cite{REF:review_ajl,REF:wph_lpp,REF:review_vw,REF:rev_gev}) the
superfluidity is manifested not by a complex scalar, as in $^4$He, but
instead by the complex vector
${\bf t}\equiv\big({\bf m}+i{\bf n}\big)/\sqrt{2}$.

This recognition, that the parametrization of $d_{\mu j}$ in terms of
$\chi$ and $\{{\bf l},{\bf m},{\bf n}\}$ is not unique, is crucial
\cite{REF:fiber}.  At any point in (real) space one may choose, e.g.,
$\chi=0$, and thus select a particular $\{{\bf l},{\bf m},{\bf n}\}$.
A choice for $\chi$ and the pair $\{{\bf m},{\bf n}\}$  at some other
point may, if one wishes, then be determined as follows:  given a path
between the two points, the orientation of $\{{\bf m},{\bf n}\}$ at the
second point is obtained by the parallel transport (defined below) of
the pair over the (curved) surface of orientations of ${\bf l}$ that
are encountered along the real-space path.  However, owing to the
nonintegrability of this parallel-transport law, the orientation of
$\{{\bf m},{\bf n}\}$ cannot be expressed as a single-valued function
of ${\bf l}$ over the entire ${\bf l}$-sphere.  Thus, if the
(real-space) path is closed then, although ${\bf l}$ returns to its
original value, ${\bf m}$ and ${\bf n}$ and therefore $\chi$ will, in
general, not (i.e., there can be anholonomy).

The parallel transport of triads over the ${\bf l}$-sphere is
accomplished via the connection
\begin{equation}
{\rm Im}\,{t}_{j}^{\ast}\,
\nabla_{\bf l}\,{t}_{j}={\bf 0},
\label{EQ:connection}
\end{equation}
the line integral of which turns out to be the solid angle swept out by
${\bf l}$ as the path is traversed \cite{REF:mvb_qpfyl}, and gives the
anholonomy mentioned above.

\noindent{\it Circulation and geometry\/}:
Having discussed geometric issues \cite{REF:stone_ab} associated with
the superfluid $^3$He order parameter matrix, we now turn to the
relationship between these issues and the circulation $\kappa$.  The
local superfluid velocity ${{\bf v}^{\rm(s)}}$ \cite{REF:momentum} can
be expressed in terms of the orbital triad
$\{{\bf l},{\bf m},{\bf n}\}$
(or, equivalently, the complex vector ${\bf t}$) via
\FL
\begin{equation}
2M{{\bf v}^{\rm(s)}}/\hbar=
m_{j}
\nabla_{\bf r}\,n_{j}+\nabla_{\bf r}\,\chi
=
{\rm Im}\,
{t}_{j}^{\ast}\,
\nabla_{\bf r}\,{t}_{j}
+\nabla_{\bf r}\,\chi.
\label{EQ:sfv}
\end{equation}
That ${\bf v}^{\rm(s)}$ can be interpreted as the superfluid velocity
follows from its behavior under galilean transformations: if
${\bf t}\rightarrow{\bf t}^{\prime}=
{\bf t}\exp\big(2iM\,\delta{\bf v}\cdot{\bf r}/\hbar\big)$ then
${{\bf v}^{\rm(s)}}\rightarrow{{\bf v}^{\rm(s)}}+\delta{\bf v}$.
A local rephasing of $\chi$
and rotating of $\{{\bf m},{\bf n}\}$ (such that $d_{\mu j}$ is
unchanged) amounts to a gauge transformation, and thus leaves
${{\bf v}^{\rm(s)}}$ unchanged.  The velocity ${{\bf v}^{\rm(s)}}$ has
the form of a gauge potential, like the London current in a
superconductor \cite{REF:jrs_tos}.  In the present context, this gauge
potential is referred to as the Berry connection.  As ${\bf d}$ is a
real vector in the A-phase of $^3$He, it produces no contribution to
${{\bf v}^{\rm(s)}}$, even if ${\bf d}$ is inhomogeneous.

Now, $\kappa$ is the line integral around a closed path
${\cal C}_{r}$ in real space:
$\kappa\equiv\oint_{{\cal C}_{r}}d{\bf r}\cdot{{\bf v}^{\rm(s)}}$.
By using Eq.~(\ref{EQ:sfv}) $\kappa$ can be expressed as
\begin{mathletters}
\begin{eqnarray}
\kappa
&=&
({\hbar}/{2M})
\oint_{{\cal C}_{r}}
d{\bf r}
\cdot
\left\{
{\rm Im}\,
{t}_{j}^{\ast}\,
\nabla_{\bf r}\,{t}_{j}+
\nabla_{\bf r}\,\chi
\right\}
\\
&=&
({\hbar}/{2M})
\oint_{{\cal C}_{l}}d{\bf l}\cdot
\big\{
{\rm Im}\,
{t}_{j}^{\ast}\,
\nabla_{\bf l}\,{t}_{j}+
\nabla_{\bf l}\,\chi
\big\},
\label{EQ:line_form}
\end{eqnarray}
\end{mathletters}where ${\cal C}_{l}$ is the contour on the
${\bf l}$-sphere traced out by the orbital vector ${\bf l}$ as the path
${\cal C}_{r}$ is circumnavigated.  By adopting the parallel-transport
connection, Eq.~(\ref{EQ:connection}), we see that the penultimate term
in Eq.~(\ref{EQ:line_form}) vanishes, so that $\kappa$ is determined by
the anholonomy $\chi({\cal C}_{l})$ in $\chi$, which is necessary to
compensate for the anholonomy in $\{{\bf m},{\bf n}\}$ induced by the
parallel-transport connection:
\begin{equation}
\kappa=
({\hbar}/{2M})
\chi({\cal C}_{l}).
\label{EQ:equality}
\end{equation}
Hence, we see that {\it the superfluid circulation is given by the
anholonomy in the parallel transport of the order parameter\/},
and that $\kappa$ is the flux associated with the Berry connection.
%
The computation of $\chi({\cal C}_{l})$ \cite{REF:mvb_qpfyl} gives
\begin{equation}
\chi({\cal C}_{l})=
\Omega({\cal C}_{l})+\!2\pi p,
\label{EQ:anholonomy}
\end{equation}
where $\Omega({\cal C}_{l})$ is the solid angle subtended by
${\cal C}_{l}$ at the center of the ${\bf l}$-sphere and $p$
is an integer.  Although $\Omega({\cal C}_{l})$ is defined
only modulo $4\pi$, the term $2\pi p$ (including odd $p$)
results in the standard circulation quantum.

For a nonsingular order parameter matrix in a simply-connected sample,
Eq.~(\ref{EQ:anholonomy}) is of course the well-known Ho circulation
theorem, obtained by Ho in a different manner \cite{REF:tlh}.  The
parallel-transport approach adopted here yields an extension to
multiply-connected samples, first conjectured in
Ref.~\cite{REF:heh_jrh}.  The extra term, $2p\pi$, allows for
singularities, which render the sample multiply-connected.  We
emphasize the anholonomic essence of the circulation, even in
simply-connected samples.  We note that Eq.~(\ref{EQ:anholonomy}) holds
whether or not the distributed vorticity
$\nabla\times{\bf v}^{\rm(s)}$ of
%
	the superflow,
%
which is given by the Mermin-Ho \cite{REF:ndm_tlh} equation
${\nabla}_{\bf r}\times{{\bf v}^{\rm(s)}}
=({\hbar}/{4M})\epsilon_{ijk}\,l_{i}\,
\big({\nabla}_{\bf r}\,l_{j}\times{\nabla}_{\bf r}\,l_{k}\big)$,
is zero.

We now come to our primary observation.  In a multiply-connected sample
from which distributed vorticity is absent the circulation $\kappa$ is
{\it quantized with an offset\/}, in striking contrast with superfluid
$^4$He.  As we elucidate below, this amounts to Aharonov-Bohm physics
stemming from geometry.  The offset is determined by the solid angle
$\Omega({\cal C}_{l})$ via Eqs.~(\ref{EQ:equality}) and
(\ref{EQ:anholonomy}).  Furthermore, the offset of $\kappa$ results in
truly stable (i.e., equilibrium) persistent flow.  For a system with
radius of order $1\,{\rm mm}$ the speed of this equilibrium flow will
be of order $3\times 10^{-3}\,{\rm mm\,s}^{-1}$.  It is desirable to
exclude distributed vorticity because whenever the area between nearby
(real-space) paths is penetrated by distributed vorticity, a continuous
variation of a path will lead to a continuous variation of $\kappa$.
(The presence of a small amount of distributed vorticity will smear the
sharpness of the quantization, but not destroy it
altogether.)\thinspace\ By contrast, if the sample is simply-connected
and there is distributed vorticity then $\kappa$ is not quantized
(i.e., a continuous variation in the path leads to a continuous
variation in $\kappa$).  However, contrary to the standard view,
quantization is absent not only in the bulk but also at surfaces, owing
to their geometry. For example, for surfaces with nonzero gaussian
curvature (such as hyperboloids) there is no quantization.
(If the sample is simply-connected and there is no distributed vorticity
then $\kappa$ is zero.)

What is the connection between Aharonov-Bohm physics and offset
circulation-quantization?  Due to the structure of the $^3$He-A order
parameter, the distributed vorticity of the superflow velocity field
${\bf v}^{\rm(s)}$ is, in general, nonzero.  Hence, in a generic sample
a continuum of values of the circulation can be found. If the
distributed vorticity vanishes and the sample is simply-connected then
for all circulation-paths the circulation is zero.  However, if the
sample is multiply-connected then, even if the distributed vorticity
vanishes, the circulation need not be zero.  This is Aharonov-Bohm
physics \cite{REF:meso_pc}.  A close analogy with Aharonov-Bohm--type
phenomena from mesoscopic physics is evident.  Consider a thin-walled
normal metal cylinder in a homogeneous axial magnetic
field\cite{REF:AAS}.  If the wall is sufficiently thin that a
negligible amount of magnetic flux penetrates the metal itself then the
conductance of the cylinder oscillates with the magnetic flux threading
the interior of the cylinder because all topologically-equivalent
Feynman (electron) paths enclose identical magnetic flux.  If the wall
is thicker, so that a non-negligible amount of flux penetrates the
metal, then the amplitude of the oscillations is diminished, as
topologically-equivalent paths now enclose differing fluxes. In
$^3$He-A, the superfluid velocity plays the role of the magnetic vector
potential, and the distributed vorticity of the superflow plays the
role of the magnetic field.  A quantization offset, the analogue of a
threading flux, requires a nontrivial $\Omega({\cal C}_{l})$, which can
be maintained, e.g., by a surface or an external field that induces an
appropriate texture in ${\bf l}$.

\noindent{\it Illustrative settings\/}:
We now illustrate the general ideas presented above by discussing two
novel and experimentally feasible settings in which offsets may be
observed.  We note that several authors have considered a variety of
textures for $^3$He-A, with and without singularities; see, e.g.,
Refs.~\cite{REF:review_vw,REF:ndm_tlh,REF:adgr,REF:textures,REF:Arai}.
(Equilibrium implications of geometric phases in the mesoscopic context
have been discussed in Ref.~\cite{REF:LGB}.)\thinspace\  In the first
setting, a sample of $^3$He-A is contained in the cavity between a pair
of long, coaxial, truncated cones of vertical angle $\alpha$.  The
texture is maintained by surface orientation effects:  the vector
${\bf l}$ tends to be oriented perpendicular to walls \cite{REF:adgr}.
We suppose that the separation between the cones is sufficiently small
that the texture \cite{REF:general} adopts the form shown in
Fig.~\ref{FIG:figone} and defined in the caption. Then, from
Eq.~(\ref{EQ:sfv}) we see that
\begin{equation}
{{\bf v}^{\rm(s)}}=({\hbar/{2M\rho}}){\bf e}_{\varphi}\sin\alpha,
\end{equation}
and ${\nabla}\times{{\bf v}^{\rm(s)}}={\bf 0}$.  Moreover, for this
conical texture ${\nabla}\times{\bf l}={\bf 0}$  (see
\cite{REF:momentum}).  The offset of $\kappa$ for any path surrounding
the axis once is given by
$({\hbar}/{2M})\Omega\left({\cal C}_{l}\right)$,
where $\Omega\left({\cal C}_{l}\right)$ is the solid angle subtended by
the normal to the conical surfaces.  This offset induces a persistent
equilibrium current.

A persistent ground-state current in a conical container can be
understood in terms of angular momentum conservation.  To see this,
consider a cylinder ($\alpha=0$).  Then
$\Omega\left({\cal C}_{l}\right)=2\pi$,
the axial component of net relative orbital
angular momentum vanishes, and $\kappa=0$.  Now imagine varying
$\alpha$ so that the cylinder becomes a cone, all the while maintaining
cylindrical symmetry.  Then, no axial torque acts during this
variation, and the increase in the axial component of the net relative
orbital angular momentum is compensated by the appearance of
center-of-mass orbital angular momentum and, hence, circulation.


We remark that although the truncated cone and the cylinder are
topologically equivalent they are, of course, geometrically
inequivalent, this inequivalence being characterized by the vertical
angle $\alpha$.   We hope that the complication arising from the
truncations of the cones will, at least in the case of long cones, not
be too severe.

The second setting concerns a magnetic-field--induced spin-texture in
the $A_1$ phase of superfluid $^3$He.  As this phase is spin-polarized,
so that only spin-projection $+1$ Cooper pairs form the condensate,
mass supercurrents imply spin supercurrents.  The order parameter
matrix has the form
$d_{\mu j}=
\frac{1}{2}
\Delta_{\uparrow\uparrow}\,
\big(d_{\mu}+
ie_{\mu}\big)
\big(m_{j}+in_{j}\big){\rm e}^{i\chi}$.
Here, $\Delta_{\uparrow\uparrow}$ is the (real) gap parameter,
$\{{\bf f},{\bf d},{\bf e}\}$ is an orthonormal triad of vectors
associated with the spin angular momentum sector, and $\chi$ is a
phase. In the $A_1$ phase, the orbital and spin sectors are
identical in structure, and the superfluid velocity
${\bf v}^{\uparrow}$ is given by
\begin{equation}
{\bf v}^{\uparrow}=
({\hbar}/{2M})
\big\{
{\rm Im}\,
\big(
t_{j}^{*}\,{\nabla}_{\bf r}\,t_{j}
+\tau_{j}^{*}\,{\nabla}_{\bf r}\,\tau_{j}
\big)
+\nabla_{\bf r}\,\chi
\big\},
\end{equation}
where ${\tau}\equiv\big({\bf d}+i{\bf e}\big)/\sqrt{2}$.
The result for $\kappa$ then reads:
\begin{equation}
\kappa=
({\hbar}/{2M})
\big\{
\Omega_{l}\big({\cal C}_{l}\big)+
\Omega_{f}\big({\cal C}_{f}\big)+
2\pi p
\big\}.
\end{equation}
Both ${\bf l}$- and ${\bf f}$-textures may lead to the offset of
$\kappa$ and to a persistent equilibrium current.  Consider a sample
contained in the cavity between a pair of coaxial cylinders.  A static,
uniform magnetic field ${\bf B}_{\rm m}$ is applied to stabilize the
$A_{1}$-phase (see \cite{REF:review_ajl,REF:wph_lpp,REF:review_vw}).
In this phase, the spin-orbit interaction \cite{REF:review_ajl} tends
to align (or anti-align) ${\bf f}$ and ${\bf l}$.  If ${\bf l}$ is
strongly anchored by the surface, as it would be in a sufficiently
narrow cavity, then for ${\bf B}_{\rm m}={B}_{\rm m}{\bf e}_{z}$ we
have ${\bf f}={\bf e}_{\rho}\cos\beta-{\bf e}_{z}\sin\beta$, where
$\tan\beta\equiv{B}_{\rm m}/{B}_{\rm so}$ and ${B}_{\rm so}
(\approx 2.8\,{\rm mT})$ is the magnitude of the effective magnetic
field \cite{REF:review_ajl} due to the spin-orbit interaction
\cite{REF:YLG}.  The solid angle subtended by ${\bf f}$ is then given
by $\Omega_{f}=2\pi\big(1-\sin\beta\big)$.  This solid angle and,
correspondingly, the persistent current, can be tuned by varying of
$B_{\rm m}$.  By tilting ${\bf B}_{\rm m}$ away from the cylinder axis,
$\Omega_{f}$ is reduced.  In the present setting, the sense of the
current is determined by the magnetic field, which explicitly breaks
time-reversal symmetry.  In the first setting, by contrast,
time-reversal symmetry is spontaneously broken (i.e., ${\bf l}$ may
point inward or outward), the sense of the current being determined by
the exhibited choice.  Spin textures may also be controlled via the
interplay between ${\bf B}_{\rm m}$ and a second field (produced, e.g.,
by a axial current-carrying wire) that is inhomogeneous on the
length-scale of the sample.

\noindent{\it Flux-quantization\/}:
We now turn to implications of geometric phases in the context of
flux-quantization in superconductors with nonscalar order parameters.
In a superconductor that exhibits a spin-triplet pairing state
\cite{REF:LPG} and a corresponding order parameter matrix
$\Delta_{\mu j}$ the current is given by
\begin{equation}
{\bf j}=
\big(2e\hbar/M\big)\,
{\rm Im}\,
\Delta_{\mu j}^{\ast}
\left(-i{\nabla}+2e{\bf A}/\hbar\right)
\Delta_{\mu j},
\label{EQ:sc}
\end{equation}
where $e$ is the electronic charge.  Following Ref.~\cite{REF:byandya},
we consider a ring of this superconductor, integrate over a path deep
inside the ring (where the current is zero), and obtain the result that
the enclosed magnetic flux is quantized with an offset.  Similarly, we
anticipate a phase-shift in Little-Parks oscillations
\cite{REF:LandP}.  Moreover, it would be interesting to test these
ideas in the context of high-temperature superconductivity. Related
effects may occur in the context of neutron star physics (see, e.g.,
Ref.~\cite{REF:review_vw}, sec.~6.2.5).

\noindent{\it Conclusions\/}:
We have considered implications of geometric phases in the context of
nonscalar superfluidity and superconductivity.  We have shown that the
geometry of the order parameter can have fascinating quantum-mechanical
ramifications, especially in the realm of circulation- and
flux-quantization.  In particular, we have shown that, in the absence
of distributed vorticity, the possible values of the circulation are
determined solely by the anholonomy in the parallel transport of the
order-parameter triad along the circulation-path.  We have also
discussed certain experimental settings in which the effects proposed
here might be observed.  Thus, we have demonstrated that, even when
irrotational, superflow in $^3$He-A is strikingly different from that
in $^4$He and $^3$He-B.  The present work may have implications for the
experiments reported in Ref.~\cite{REF:Zieve}

\smallskip
It is a pleasure to thank I.~Aleiner, J.~Davis, D.~Osheroff,
M.~Zapotocky and, especially, A.~J.~Leggett and M.~Stone for useful
discussions.
This work was supported by NSF via ECS91-08300 (YLG), DMR91-57018 and
DMR94-24511 (PMG) and NSERC Canada (DL).

\figure{
a)~Concentric truncated cones of vertical angle $\alpha$ with
superfluid $^3$He-A in the cavity. The texture is given by
$\chi=0$ and (for cylindrical polar
coordinates $\{\rho,\varphi,z\}$)
$\{ {\bf l},{\bf m},{\bf n}\}=
 \{-{\bf e}_{\rho}\cos\alpha+{\bf e}_{z}\sin\alpha,
    {\bf e}_{\rho}\sin\alpha+{\bf e}_{z}\cos\alpha,
    {\bf e}_{\varphi}\}$.
The texture has ${\bf l}$ pointing from outer to inner cone
\cite{REF:alternate}.
${\cal C}_{r}$ is a path along which $\kappa$ is computed.
b)~${\bf l}$-sphere with tangent plane containing
$\{{\bf m},{\bf n}\}$, and ${\cal C}_{l}$ is the image of ${\cal C}_{r}$.
\label{FIG:figone}}

\begin{references}
\bibitem[\ast]{YGnote}
E-mail address: {\tt lyanda@ceg.uiuc.edu\/}
\bibitem[\dag]{PGnote}
E-mail address: {\tt goldbart@uiuc.edu\/}
\bibitem[\ddag]{DLnote}
E-mail address: {\tt dloss@sfu.ca\/}
\bibitem{REF:lo_rpf_rd}
F.\ London, Phys.\ Rev.\ 74, 562 (1948);
R.\ Doll, M.\ N{\"a}bauer,
Phys.\ Rev.\ Lett.\ 7, 51 (1961);
B.\ Deaver, W.\ Fairbank,
ibid., 7, 43 (1961).
\bibitem{REF:byandya}
N.\ Byers, C.\ N.\ Yang,
Phys.\ Rev.\ Lett.\ 7, 46 (1961).
\bibitem{REF:review_ajl}
A.\ J.\ Leggett,
Rev.\ Mod.\ Phys.\ 47, 331 (1975).
\bibitem{REF:wph_lpp}
{\sl Helium Three\/}, edited by
W.\ P.\ Halperin, L.\ P.\ Pitaevskii
(North-Holland, Amsterdam, 1990).
\bibitem{REF:review_vw}
D.\ Vollhardt, P.\ W{\"o}lfle,
{\sl The Superfluid Phases of Helium Three\/}
(Taylor and Francis, London, 1990).
\bibitem{REF:aha_boh}
Y.\ Aharonov, D.\ Bohm,
Phys.\ Rev.\ 115, 485 (1959).
\bibitem{REF:as_fw}
For a review, see
A.\ Shapere, F.\ Wilczek,
{\sl Geometric Phases in Physics\/}
(World Scientific, Singapore, 1989).
\bibitem{REF:LandP}
W.\ A.\ Little, R.\ D.\ Parks,
Phys.\ Rev.\ Lett.\ 9, 9 (1962).
\bibitem{REF:mvb_qap}
M.\ V.\ Berry,
Proc.\ R.\ Soc.\ Lon.,
Ser.~A 392, 45 (1984).
\bibitem{REF:mvb_qpfyl}
See M.\ V.\ Berry, in Ref.~\cite{REF:as_fw}, p.~8 et seq.
\bibitem{REF:rev_gev}
G.\ E.\ Volovik,
{\sl Exotic Properties of Superfluid $^3$He\/}
(World Scientific, Singapore, 1992).
\bibitem{REF:fiber}
One may say that at each point in the base space (the ${\bf l}$-sphere)
there is a fiber of orientations of the pair $\{{\bf m},{\bf n}\}$ in
the plane tangent to the point ${\bf l}$. Together, the orientations
and the ${\bf l}$-sphere make a frame-bundle. See B.\ Simon,
Phys.\ Rev.\ Lett.\ 51, 2167 (1983), reprinted in Ref.~\cite{REF:as_fw};
and, e.g., Ref.~\cite{REF:as_fw}, Chap.~3.
\bibitem{REF:stone_ab}
For other implications of geometric phases in superfluid $^3$He,
see, e.g.,
W.\ E.\ Goff, F.\ Gaitan, M.\ Stone,
Phys.\ Lett.\ A136, 433 (1989);
A.\ Garg et al.,
Ann.\ Phys.\ (NY) 173, 149 (1987).
The Aharonov-Casher effect is considered in
A.\ Balatsky, B.\ L.\ Altshuler,
Phys.\ Rev.\ Lett.\ 70, 1678 (1993).
\bibitem{REF:momentum}
The superfluid current density in $^3$He-A contains two terms
proportional to ${\nabla}\times{\bf l}$ as well as the term
determined by the superfluid velocity ${{\bf v}^{\rm(s)}}$.
If ${\nabla}\times{\bf l}={\bf 0}$ then the superfluid current
density is due solely to ${{\bf v}^{\rm(s)}}$.
\bibitem{REF:jrs_tos}
See, e.g.,
J.\ R.\ Schrieffer,
{\it Theory of Superconductivity\/}
(Addison-Wesley, 1983), pp.~10-17.
\bibitem{REF:tlh}
T.-L.\ Ho,
Phys.\ Rev.\ B 18, 1144 (1978).
\bibitem{REF:heh_jrh}
H.\ E.\ Hall, J.\ R.\ Hook,
{\it Prog.\ in Low Temp.\ Physics\/},
Vol.~9, D.\ Brewer (ed.) (North-Holland, 1986) p.~143.
\bibitem{REF:ndm_tlh}
N.\ D.\ Mermin, T.-L.\ Ho,
Phys.\ Rev.\ Lett.\ 36, 594 (1976).
\bibitem{REF:meso_pc}
Superflow quantized with an offset may be viewed as a macroscopic
analog of the persistent current caused by magnetic flux in a
mesoscopic metallic ring.
\bibitem{REF:AAS}
B.\ L.\ Altshuler, A.\ G.\ Aronov, B.\ Z.\ Spivak,
Zh.\ Eksp.\ Teor.\ Fiz.\
33, 101 (1981)
[JETP Lett.\ 33, 94 (1981)].
\bibitem{REF:adgr}
V.\ Ambegaokar et al.,
Phys.\ Rev.\ A 9, 2676 (1974).
\bibitem{REF:textures}
P.\ Anderson, W.\ Brinkman,
in {\sl Helium Liquids\/},
J.\ Armitage, I.\ Farquhar (eds.),
(Academic Press, London, 1974), p.~315;
P.\ Anderson, G.\ Toulouse, Phys.\ Rev.\ Lett.\ 38, 508 (1977);
A.\ Fetter, Phys.\ Rev.\ B 15, 1350 (1977);
K.\ Maki, X.\ Zotos, Phys.\ Rev.\ B 31, 177 (1985);
P.\ Muzikar, J.\ Low Temp.\ Phys.\ {\bf 46\/}, 533 (1982).
\bibitem{REF:Arai}
T.\ Arai, T.\ Soda, Prog.\ Theor.\ Phys.\ {\bf 69\/}, 699 (1983).
\bibitem{REF:LGB}
D.\ Loss et al.,
Phys.\ Rev.\ Lett.\ 65, 1655 (1990);
D.\ Loss, P.\ M.\ Goldbart, Phys.\ Rev.\ B 45, 13544 (1992).
\bibitem{REF:general}
For more general textures (e.g., for {\it generalized cones\/}, i.e.,
surfaces constructed by sweeping a straight generatrix fixed at one
point around a closed path) the geometric phase viewpoint provides an
efficient computational tool.
\bibitem{REF:alternate}
Alternatively, there is a degenerate texture in which ${\bf l}$ points
from the inner to the outer cone. For narrow cavities, ${\bf l}$ is
perpendicular to the surfaces throughout the cavity.  The favorability
of similar textures between concentric cylinders has been considered in
Ref.~\cite{REF:Arai}.  Excited states containing domain walls between
textures with opposite ${\bf l}$ are expected to be rather costly in
energy.
\bibitem{REF:YLG}
Spin-orbit geometric phases in mesoscopic rings are discussed in
A.\ Aronov, Y.\ Lyanda-Geller,
Phys.\ Rev.\ Lett.\ 70, 343 (1993);
Y.\ Lyanda-Geller,
ibid., 71, 657 (1993).
\bibitem{REF:LPG}
In the cubic heavy-fermion superconductors
(such as ${\rm U\,Be}_{13}$) an order parameter triad is symmetry-allowed.
See, e.g., L.\ P.\ Gor'kov,
Sov.\ Sci.\ Rev.\ A Phys.\ 9, 1-116 (1987), sec.~6.
\bibitem{REF:Zieve}
R.\ J.\ Zieve, Yu.\ Mukharsky, J.\ D.\ Close, J.\ C.\ Davis,
R.\ E.\ Packard,
J.\ Low Temp.\ Phys.\ 89, 47 (1992); we thank
J.\ C.\ Davis for drawing our attention to this paper.
\end{references}
\end{document}